\documentclass[a4paper,12pt]{article}

\usepackage{amsmath}
\usepackage{amsfonts}
\usepackage{youngtab} 

\newcommand{\del}{{\delta}}

\newcommand{\kap}{{\kappa}}
\newcommand{\la}{{\lambda}}
\newcommand{\La}{{\Lambda}}
\newcommand{\om}{{\omega}}
\newcommand{\sig}{{\sigma}}

\newcommand{\half}{{\textstyle{\frac{1}{2}}}}
\newcommand{\nn}{{\nonumber}}

\newcommand{\eqand}{{\quad \mathrm{and} \quad}}

\newcommand{\proof}{\paragraph*{Proof:}}
\newtheorem{lemma}{Lemma}

\newcommand{\lb}{{\ell}}
\newcommand{\nb}{{n}}
\newcommand{\mb}[1]{{m_{(#1)}}}

\newcommand{\M}[1]{{\stackrel{#1}{M}}}  

\newcommand{\Om}{\Omega} 
\newcommand{\Ps}{\Psi}   
\newcommand{\vphi}{\varphi} 
\newcommand{\vphip}{\varphi'} 
\newcommand{\Phis}{\Phi^\mathrm{S}} 
\newcommand{\Phia}{\Phi^\mathrm{A}} 
\newcommand{\Tf}{T^\Phi}
\newcommand{\Tp}{T^\Psi}
\newcommand{\To}{T^\Om}

\newcommand{\taup}{\tau'}
\newcommand{\rhop}{\rho'}

\newcommand{\kappap}{\kappa'}

\DeclareTextFontCommand{\textwasy}{\wasyfamily}
\def \wasyfamily{\fontencoding{U}\fontfamily{wasy}\selectfont}
\def \thorn{{\wasyfamily\char105}}
\DeclareTextCommand{\dh}{OT1}{{\wasyfamily\char107}}
\newcommand{\tho}{{\textrm\thorn}}
\newcommand{\eth}{{\textrm{\dh}}}
\newcommand{\thop}{\tho'}

\numberwithin{equation}{section}

\title{Perturbations of higher-dimensional spacetimes}
\author{Mark Durkee and Harvey S. Reall\\
        {\small\it DAMTP, Centre for Mathematical Sciences, University of Cambridge,}\\
        {\small\it Wilberforce Road, Cambridge, CB3 0WA, United Kingdom}\\
        {\small\tt M.N.Durkee@damtp.cam.ac.uk, H.S.Reall@damtp.cam.ac.uk}}

\begin{document}
\maketitle
\begin{abstract}
We discuss linearized gravitational perturbations of higher dimensional spacetimes. For algebraically special spacetimes (e.g.\ Myers-Perry black holes), we show that there exist local gauge invariant quantities linear in the metric perturbation. These are the higher dimensional generalizations of the 4d Newman-Penrose scalars that (in an algebraically special vacuum spacetime) satisfy decoupled equations of motion. We show that decoupling occurs in more than four dimensions if, and only if, the spacetime admits a null geodesic congruence with vanishing expansion, rotation and shear. Decoupling of electromagnetic perturbations occurs under the same conditions. Although these conditions are not satisfied in black hole spacetimes, they are satisfied in the near-horizon geometry of an extreme black hole.
\end{abstract}


\section{Introduction}

The last decade has seen increasing interest in the study of General Relativity in $d>4$ spacetime dimensions.  Black hole solutions are of particular interest \cite{ER:2008}, and the classical stability of such solutions is of obvious importance. It has been established that the Schwarzschild solution is stable against linearized gravitational perturbations for all $d>4$ \cite{Ishibashi:2003}.  However, heuristic arguments \cite{Emparan:2003sy} suggest that Myers-Perry black holes \cite{mp} might be unstable for sufficiently large angular momentum.  

The prediction of Ref.\ \cite{Emparan:2003sy}  has been confirmed by a linearized stability analysis for certain Myers-Perry black holes with enhanced symmetry. Refs.\ \cite{Dias:2009iu,Dias:2010maa} considered the case of a singly-spinning MP black hole, i.e., one with a single non-vanishing angular momentum. It was shown that, for a critical value of the angular momentum (for fixed mass), there exists a stationary gravitational perturbation. It was argued that this corresponds to the threshold of instability, i.e., black holes with larger angular momentum are unstable. Ref.\ \cite{Dias:2010eu} considered the most symmetrical case, a MP black hole with equal angular momenta in an odd number of dimensions, and demonstrated the existence of gravitational perturbations growing exponentially with time, for sufficiently large angular momentum. Instabilities of singly-spinning MP black holes have also been found via nonlinear numerical evolution of a perturbed black hole \cite{Shibata:2009ad}. 

In spite of this recent progress, a study of the general case seems hopelessly difficult owing to the complexity of the linearized equations of motion, or the difficulty of performing numerical evolution of solutions with a large number of free parameters.

So far, studies of linearized perturbations of higher-dimensional black holes have exploited {\it isometries} of black hole spacetimes, e.g. spherical symmetry of the Schwarzschild solution, or enhancement of symmetry of the Myers-Perry solution when some of the angular momenta coincide. However, in 4 dimensions, there is an alternative approach, due to Teukolsky \cite{Teukolsky:1972,Teukolsky:1973}, which exploits the algebraically special nature of black hole solutions. It is this approach which renders tractable the study of perturbations of the Kerr solution.

The Teukolsky approach is as follows.  Consider a 4d spacetime, and let $(\ell,n,m,\bar{m})$ be a null tetrad. The Weyl tensor is encoded in the Newman-Penrose scalars $\Psi_0,\ldots, \Psi_4$.  Now, consider a linearized perturbation of such a spacetime. Let $\Psi_A^{(0)}$ denote the unperturbed value of $\Psi_A$, and let $\Psi_A^{(1)}$ denote the perturbation. In general, there is gauge freedom corresponding to the possibility of infinitesimal coordinate transformations and infinitesimal changes of tetrad.  However, it can be shown that $\Psi_0^{(1)}$ is gauge invariant if (and only if) $\ell$ is a repeated principal null direction of the background spacetime.  Therefore, for perturbations of algebraically special spacetimes, there exists a local, gauge-invariant quantity, linear in the metric perturbation. It is natural to exploit this fact when studying perturbations of such spacetimes.

In a general spacetime, the linearized equations of motion will lead to coupled equations for the quantities $\Psi_A^{(1)}$. Remarkably, in an algebraically special vacuum spacetime, Teukolsky showed that one can decouple these equations to obtain a single, second order, wave equation for $\Psi_0^{(1)}$.  If the background is type D (for example, the Kerr metric), i.e., if $\ell$ and $n$ both are repeated principal null directions, then $\Psi^{(1)}_0$ and $\Psi^{(1)}_4$ both are gauge invariant and both satisfy decoupled equations of motion. The solutions of these two equations are related by identities, and knowledge of either one is sufficient to reconstruct the metric perturbation.

The goal of the present paper is to investigate whether, and how, these gauge invariance and decoupling properties extend to perturbations of higher-dimensional spacetimes. It is clear from the outset that, for $d>4$, we should not expect to obtain a full description of gravitational perturbations solely in terms of a single complex scalar analagous to $\Psi_0^{(1)}$ (or $\Psi_4^{(1)}$). This is because the gravitational field in $d>4$ dimensions has more physical degrees of freedom than a complex scalar. 

The appropriate generalization of $\Psi_0$ can be identified by appealing to the classification of the Weyl tensor in $d \ge 4$ dimensions \cite{cmpp}. This uses a basis $e_0=\ell$, $e_1=n$, $e_i=m_{(i)}$, $i=2,\ldots d-1$, where $\ell$ and $n$ are null and $m_{(i)}$ are spacelike.
The higher-dimensional analogue of $\Psi_0$ is a $(d-2) \times (d-2)$ traceless symmetric matrix $\Omega_{ij}\equiv C_{0i0j}$, where $C_{abcd}$ is the Weyl tensor. The analogue of $\Psi_4$ is another such matrix, $\Omega'_{ij} \equiv C_{1i1j}$. These quantities transform as scalars under general coordinate transformations.  Note that the number of independent components of $\Omega_{ij}$ (or $\Omega'_{ij}$) is the same as the number of physical degrees of freedom of the gravitational field.

In $d>4$ dimensions, the analogue of a principal null direction is a {\it Weyl aligned null direction} (WAND), and the analogue of a repeated principal null direction is a {\it multiple WAND} \cite{cmpp}. We say that a spacetime is algebraically special if it admits a multiple WAND. It is type D (or O) if it admits two multiple WANDs.  The MP black hole is an example of a type D spacetime \cite{mptypeD}.

Just as for $d=4$, we find that $\Omega_{ij}^{(1)}$ is invariant under infinitesimal coordinate transformations and infinitesimal changes of basis if (and only if) $\ell$ is a multiple WAND.  $\Omega_{ij}^{(1)}$ and $\Omega_{ij}^{'(1)}$ both are gauge invariant if (and only if) $\ell$ and $n$ both are multiple WANDs, that is the background is type D (or O).  This gauge invariance implies that, irrespective of decoupling, these quantities are natural objects to consider when studying gravitational perturbations of higher-dimensional algebraically special solutions.

We will study linearized gravitational perturbations of algebraically special spacetimes satisfying the vacuum Einstein equation (allowing for a cosmological constant). We analyze decoupling following the approach of Stewart and Walker \cite{stewperts}. This exploits the Geroch-Held-Penrose (GHP) formalism \cite{ghp}, a very useful approach for studying spacetimes with one or two preferred null directions, such as algebraically special spacetimes. Ref.\ \cite{higherghp} extended this formalism to $d>4$. We find that $\Omega_{ij}^{(1)}$ satisfies a decoupled equation in  an algebraically special vacuum spacetime with $d>4$ if, and only if, $\ell$ is geodesic and free of expansion, rotation and shear.  We also analyze the simpler case of a Maxwell field and find that exactly the same condition is required for decoupling  in this case.

A spacetime admitting a null geodesic congruence with vanishing expansion, rotation and shear is known as a Kundt spacetime.  It was shown in Ref.\ \cite{Ricci} that any such spacetime is algebraically special (in vacuum).  Hence our result is that decoupling occurs if, and only if, the spacetime is Kundt. This result can be contrasted with $d=4$, for which decoupling requires only that $\ell$ be geodesic and shearfree. These conditions are equivalent to $\ell$ being a repeated principal null direction (by the Goldberg-Sachs theorem), i.e., they are satisfied in any algebraically special spacetime.

In order for $\Omega_{ij}^{(1)}$ and $\Omega_{ij}^{'(1)}$  both to satisfy decoupled equations, $\ell$ and $n$ both must be geodesic with vanishing expansion, rotation and shear. We will refer to such a spacetime as {\it doubly Kundt}. A doubly Kundt spacetime must be type D (or O).

Unfortunately, black hole spacetimes are not Kundt and therefore decoupling does not occur in higher-dimensional black hole spacetimes.\footnote{There is no contradiction with the results of Ref. \cite{Ishibashi:2003} since that reference studies perturbations by exploiting the spherical symmetry of the Schwarzschild solution rather than its type D property, i.e., $\Omega_{ij}^{(1)}$ is not used to describe the perturbation. The quantities that satisfy decoupled equations are {\it non-local} in the metric perturbation.} Obviously is it disappointing that decoupling does not occur for the Myers-Perry solution.\footnote{Nevertheless, we emphasize that the quantities $\Omega_{ij}^{(1)}$, $\Omega_{ij}^{'(1)}$ should be useful in studies of Myers-Perry perturbations because of their locality and gauge invariance.} However, as we will explain at the end of this paper, the  {\it near-horizon geometries} of known extreme vacuum black hole solutions  {\it are} doubly Kundt solutions.  Therefore our decoupled equation is ideal for studying perturbations of near-horizon geometries. In a companion paper \cite{nearhorizon} we will demonstrate that, under certain circumstances, one can predict an instability of the full extreme black hole geometry by using our decoupled equation to demonstrate an instability of the near-horizon geometry.

This paper is organized as follows.  In Section \ref{sec:notation} we describe our notation, and give a brief summary of the necessary results regarding the higher-dimensional generalization of the GHP formalism from Ref.\ \cite{higherghp}.  In Section \ref{sec:gaugeinv} we investigate the existence of gauge invariant quantities, and show that $\Omega_{ij}^{(1)}$ are gauge invariant if and only if the background spacetime is algebraically special.  We then move on to consider decoupling of perturbations.  As a warm-up exercise, in Section \ref{sec:maxdecoupling} we consider the decoupling of Maxwell perturbations, as this simpler example illustrates the approach that we use in the gravitational case in Section \ref{sec:gravdecoupling}.  Finally, in Section \ref{sec:discussion} we discuss the possible applications of our results.

\section{Notation}\label{sec:notation}

We will make heavy use of the higher-dimensional generalization of the GHP formalism, which was developed in Ref. \cite{higherghp}. This section gives a brief summary of this formalism.  

We will use a basis $e_0=\lb$, $e_1=\nb$, $e_i=\mb{i}$, $i=2,\ldots d-1$, where
\begin{equation}\label{eqn:framedef}
 \lb^2=n^2=\lb \cdot \mb{i} = \nb \cdot \mb{i}=0, \qquad 
 \lb \cdot \nb = 1, \qquad 
 \mb{i} \cdot \mb{j} = \delta_{ij}.
\end{equation}
Changes of basis are described by Lorentz transformations. The Lorentz group is generated by the following transformations:
\begin{itemize}
  \item Boosts ($\lambda$ a real function)
        \begin{equation} \label{eqn:boosts}
          \lb \mapsto \la \lb,\qquad 
          \nb \mapsto \la^{-1} \nb,\qquad 
          \mb{i} \mapsto \mb{i}.
        \end{equation}
  \item Spins ($X_{ij} \in SO(d-2)$)
        \begin{equation} \label{eqn:spins}
          \lb \mapsto \lb,\qquad 
          \nb \mapsto \nb,\qquad 
          \mb{i} \mapsto X_{ij} \mb{j}.
        \end{equation}
  \item Null rotations about $\lb$ ($z_i$ some real functions)
        \begin{equation}
          \lb \mapsto \lb,\qquad 
          \nb \mapsto \nb + z_i \mb{i} - \tfrac{1}{2} z^2 \lb,\qquad 
          \mb{i} \mapsto \mb{i} - z_i \lb.
        \end{equation}
  \item Null rotations about $\nb$ ($z_i$ some real functions)
        \begin{equation}
          \lb \mapsto \lb + z_i \mb{i} - \tfrac{1}{2} z^2 \nb,\qquad 
          \nb \mapsto  \nb,\qquad 
          \mb{i} \mapsto \mb{i} - z_i \nb.
        \end{equation}
\end{itemize}
We define
\begin{equation}
  L_{ab} = \nabla_b l_a,\qquad
  N_{ab} = \nabla_b n_a,\qquad
  \M{i}_{ab} = \nabla_b m_{(i)a},
\end{equation}
and also set
\begin{equation}
  \rho_{ij} = L_{ij} ,\qquad
  \tau_{i} = L_{i1} ,\qquad
  \kap_{i} = L_{i0}.
\end{equation}
We can decompose $\rho_{ij}$ into its trace $\rho$, its symmetric traceless part $\sigma_{ij}$ and its antisymmetric part $\om_{ij}$. These represent the expansion, shear and twist of the null geodesic congruence defined by the vector field $\lb$. $\kappa_i$ vanishes if, and only if, this congruence is geodesic.

The idea of the GHP formalism is to maintain covariance with respect to boosts and spins. We say that an object $T_{i_1 \ldots i_s}$ is a {\it GHP scalar} of spin $s$ and boost weight $b$ if it transforms as a Cartesian tensor of rank $s$ under spins \eqref{eqn:spins}, and $T_{i_1 \ldots i_s} \rightarrow \lambda^b T_{i_1 \ldots i_s}$ under a boost \eqref{eqn:boosts}. The quantities $\rho_{ij},\tau_i,\kap_i$ are GHP scalars with $b=1,0,2$ respectively. Other quantities, e,g, $L_{10}$ are not GHP scalars because they transform inhomogeneously under boosts or spins.

Given any GHP scalar $T$, we denote by $T'$ the object obtained by exchanging $\lb$ and $\nb$ in its definition, so for example $\tau'_i = N_{i0}$.  This {\it priming operation} leads to a significant reduction in the number of equations that need to be displayed explicitly.

The heart of the GHP formalism is a set of derivative operators that map GHP scalars to GHP scalars. 
They act on a GHP scalar $T_{i_1 i_2...i_s}$ of spin $s$ and boost weight $b$ as:
\begin{eqnarray}
  \tho T_{i_1 i_2...i_s} 
      &\equiv & (\lb\! \cdot\! \partial ) T_{i_1 i_2...i_s} - b L_{10} T_{i_1 i_2...i_s}
                + \sum_{r=1}^s \M{k}_{i_r 0} T_{i_1...i_{r-1} k i_{r+1}...i_s},\\
  \tho' T_{i_1 i_2...i_s} 
      &\equiv & (\nb\! \cdot\! \partial ) T_{i_1 i_2...i_s} - b L_{11} T_{i_1 i_2...i_s}
                + \sum_{r=1}^s \M{k}_{i_r 1} T_{i_1...i_{r-1} k i_{r+1}...i_s},\\
  \eth_i T_{j_1 j_2...j_s} 
      &\equiv & (\mb{i}\! \cdot\! \partial )T_{j_1 j_2...j_s} - b L_{1i} T_{j_1 j_2...j_s}
                + \sum_{r=1}^s \M{k}_{j_r i} T_{j_1...j_{r-1} k j_{r+1}...j_s}.
\end{eqnarray}
The spin and boost weights of these objects are $(s,b+1)$, $(s,b-1)$ and $(s+1,b)$, respectively.

We expand the Weyl tensor $C_{abcd}$ in the frame (\ref{eqn:framedef}), and write
\begin{equation}
 \Om_{ij} \equiv C_{0i0j},  \qquad \Om'_{ij} \equiv C_{1i1j} 
\end{equation}
\begin{equation}
 \Psi_{ijk} \equiv C_{0ijk},\qquad   \Psi'_{ijk} \equiv C_{1ijk} ,\qquad
 \Psi_{i} \equiv C_{010k} ,\qquad   \Psi'_{i} \equiv C_{101i}
\end{equation}
and
\begin{equation}
 \Phi_{ij} \equiv C_{0i1j}, \qquad 
 \Phi_{ijkl} \equiv C_{ijkl} , \qquad
 \Phi \equiv C_{0101} , \qquad       \Phia_{ij} = \tfrac{1}{2} C_{01ij}.
\end{equation}
These quantities are all GHP scalars, and the notation reflects the boost weights: $\Omega$ has $b=2$, $\Psi$ has $b=1$, $\Phi$ has $b=0$, $\Psi'$ has $b=-1$ and $\Omega'$ has $b=-2$. Note the identities
\begin{equation}
    \Om_{ii} = 0 , \qquad   \Om_{ij} = \Om_{(ij)},
\end{equation}
\begin{equation}
  \Psi_{iji} = \Psi_j,\qquad   \Psi_{ijk} = \Psi_{i[jk]},\qquad  \Psi_{[ijk]} = 0
\end{equation}
and
\begin{equation}
  -\tfrac{1}{2} \Phi_{ijkj} = \Phis_{ik} \equiv \Phi_{(ik)},\qquad
  \Phia_{ij} = \Phi_{[ij]} ,\qquad \Phi = \Phi_{ii} = -\tfrac{1}{2}\Phi_{ijij}.
\end{equation}
The dynamical content of GR is encapsulated in the ``Newman-Penrose'' equations, Bianchi identity and the commutators of GHP derivatives. These are written out in Appendix \ref{app:ghpeqns} for the case of an Einstein spacetime:
\begin{equation}
 R_{\mu\nu} = \Lambda g_{\mu\nu}.
\end{equation}
We shall consider only Einstein spacetimes in this paper.

Finally, for a Maxwell field strength $F_{ab}$, we use the notation
\begin{equation}
 \vphi_i = F_{0i}, \qquad F=F_{01}, \qquad F_{ij} = F_{ij}, \qquad \vphi'_i = F_{1i}.
\end{equation}
These quantities are all GHP scalars and $\vphi_i$ has $b=1$, $F_{01}$ and $F_{ij}$ have $b=0$ and $\vphi'_i$ has $b=-1$.

\section{Gauge-invariant variables}\label{sec:gaugeinv}

We are interested in linearized perturbations of spacetimes. For a quantity $X$, we shall write $X=X^{(0)}+X^{(1)}$ where $X^{(0)}$ is the value in the background spacetime and $X^{(1)}$ is the perturbation.  Following Ref.\ \cite{stewperts}, we look to find variables that are gauge invariant under both infinitesimal coordinate transformations and infinitesimal changes of basis.

Let $X$ be a spacetime scalar. Then, under an infinitesimal coordinate transformation with parameters $\xi^\mu$, we have $ X^{(1)} \rightarrow X^{(1)} + \xi \cdot \partial X^{(0)}$. Hence $X^{(1)}$ is invariant under infinitesimal coordinate transformations if, and only if, $X^{(0)}$ is constant.

In the case of gravitational perturbations, we are interested in the scalars $\Omega_{ij}$ since these are the higher-dimensional generalization of the 4d quantity $\Psi_0$. We have
\begin{lemma}\label{lem:Ominv}
  $\Om^{(1)}_{ij}$ is a gauge invariant quantity if and only if $\lb$ is a multiple WAND of the background spacetime (or equivalently, if and only if $\Psi^{(0)}_{ijk} = 0 = \Om^{(0)}_{ij}$).
\end{lemma}
Note that this Lemma does not require any assumptions regarding the matter content of the spacetime.
\proof
First we consider infinitesimal basis transformations. Consider an infinitesimal spin of the form \eqref{eqn:spins}. If $\Omega^{(0)}_{ij}$ is non-vanishing then this will induce a change in $\Omega^{(1)}_{ij}$.  Hence we must have $\Omega^{(0)}_{ij}=0$ for $\Om^{(1)}_{ij}$ to be gauge invariant.

Next consider an infinitesimal null rotation about $\nb$. Using Ref. \cite{higherghp}, the change in $\Omega^{(1)}$ is, to linear order in the infinitesimal parameters $z^i$,  
\begin{equation}
  \Om^{(1)}_{ij}
     \mapsto \Om^{(1)}_{ij}-2 z_k (\Psi^{(0)}_{(i}\del_{j)k}+\Psi^{(0)}_{(ij)k})
\end{equation}
For invariance, we need 
\begin{equation}
  \Psi^{(0)}_{(i}\del_{j)k}+\Psi^{(0)}_{(ij)k} = 0.
\end{equation}
Taking the trace on $j$ and $k$ gives $\Psi^{(0)}_i=0$. We then use $\Psi_{ijk} = \tfrac{2}{3}(\Psi_{(ij)k} - \Psi_{(ik)j})$ to deduce that $\Psi^{(0)}_{ijk}=0$.  So we conclude that invariance of $\Omega^{(1)}_{ij}$ under infinitesimal basis transformations implies that
\begin{equation}
 \Omega^{(0)}_{ij} = \Psi^{(0)}_{ijk} = 0.
\end{equation}
It is easy to see that these conditions are both necessary and sufficient for $\Omega^{(1)}_{ij}$ to be invariant under infinitesimal basis transformations. These conditions are equivalent to the statement that $\ell$ is a multiple WAND of the background geometry. 

Finally, since $\Omega^{(0)}_{ij}=0$, it follows that $\Omega^{(1)}_{ij}$ is invariant under infinitesimal coordinate transformations. $\Box$

Similarly, $\Omega^{'(1)}_{ij}$ is gauge invariant if, and only if, $\nb$ is a multiple WAND. Hence both quantities are gauge invariant if, and only if, the spacetime is type D.\footnote{Note that further gauge invariant quantities exist for higher dimensional spacetimes satisfying additional restrictions, see Lemma \ref{lem:otherinvs} later.}

Now consider a Maxwell field. We shall consider only a test field, i.e., we neglect gravitational backreaction and treat the Maxwell field as an infinitesimal quantity that vanishes in the background. It follows that all components are  invariant to first order under infinitesimal coordinate transformations and infinitesimal basis transformations. Note that, since we are treating the Maxwell field as infinitesimal, and working to first order, there is no distinction between Maxwell theory and Maxwell theory with a Chern-Simons term.

So far we have discussed only {\it infinitesimal} basis transformations. However, sometimes one might want to consider finite transformations. For example, consider a type D spacetime. Then $\ell$ and $n$ are fixed (up to scaling) in the background by the requirement of being multiple WANDs. But there is not preferred way of choosing the spatial basis vector $\mb{i}$. Different choices are related by {\it finite} spins. $\Omega^{(1)}_{ij}$ and $\Omega^{'(1)}_{ij}$ are not invariant under finite spins. Exactly the same issue arises in 4d, where $\Psi_0^{(1)}$ and $\Psi_4^{(1)}$ pick up phases under finite spins. 

Physical quantities should not care about the choice of spatial basis vectors so such quantities must be related to GHP scalars with zero spin. For example, in an asymptotically flat 4d spacetime, the energy flux in ingoing and outgoing gravitational waves is related to the spin-0 GHP scalars $|\Psi_0^{(1)}|^2$ and $|\Psi_4^{(1)}|^2$, respectively (for appropriate choices of $\ell$ and $n$, see \cite{Teukolsky:1973}). For $d>4$, the analogous quantities are $\Omega^{(1)}_{ij} \Omega^{(1)}_{ij}$ and $\Omega^{'(1)}_{ij} \Omega^{'(1)}_{ij}$. We can also define additional invariant quantities such as $\Phi_{ij}^{(0)} \Omega^{(1)}_{ij}$.\footnote{In 4d this quantity vanishes because $\Phi^S_{ij} = \tfrac{1}{2} \Phi \delta_{ij}$.}

\section{Decoupling of electromagnetic perturbations}\label{sec:maxdecoupling}

\subsection{Main Result}

The highest boost weight components of the Maxwell field are denoted $\vphi_i$.
In 4d, the quantity analogous to $\vphi_i$, satisfies a decoupled equation of motion in an algebraically special background. We shall investigate the conditions under which $\vphi_i$ satisfies a decoupled equation of motion in $d>4$ dimensions. The motivation for doing this is mainly that the Maxwell field illustrates the arguments that we shall also employ in the gravitational case, but the equations are considerably simpler.

In this section, we show how, in a particular class of background Einstein spacetimes, we can construct decoupled 2nd order differential equations for a Maxwell test field.  We show that this decoupling is possible if and only if the background spacetime is Kundt, that is it admits a geodesic null vector field that is not shearing, twisting or expanding.

We will show that the dynamics of a Maxwell test field on the background of a Kundt spacetime can be described by the following equation:
\begin{equation}\label{eqn:maxpert}
    \left(2\tho'\tho + \eth_j\eth_j + \rho'\tho -4\tau_j\eth_j 
                                   + \Phi-\tfrac{2d-3}{d-1}\La\right)\vphi_i 
             + (- 2\tau_i\eth_j + 2\tau_j \eth_i + 2\Phis_{ij} + 4\Phia_{ij})\vphi_j = 0.  
\end{equation}
We also show that analogous decoupled equations cannot be constructed for spacetimes that are not Kundt, and discuss briefly whether any alternative progress can be made.

It is interesting to compare this to the equation of motion for a massive scalar field $\phi$:
\begin{equation}
  (\nabla_\mu \nabla^\mu - \mu^2) \phi = 0.
\end{equation}
When written out in GHP form in a general background, this equation is
\begin{equation} \label{eqn:scalarperts}
  (2\tho'\tho + \eth_i\eth_i + \rho'\tho  -2\tau_i\eth_i + \rho\tho'  - \mu^2)\phi = 0.
\end{equation}
To compare this with the decoupled Maxwell equation, one must specialize to a Kundt spacetime, for which $\rho=0$. Note that $\tau'_i$ does not appear in either equation.

\subsection{Derivation of main result}

The Maxwell equations for a 2-form field strength are \cite{higherghp}:
\begin{eqnarray}
  \eth_i \vphi_i + \tho F\label{max1}
         &=& \tau'_i \vphi_i + \rho_{ij} F_{ij} - \rho F - \kap_i \vphi'_i \\
  2 \eth_{[i} \vphi_{j]} - \tho F_{ij} \label{max2}
         &=& 2\tau'_{[i} \vphi_{j]} + 2F \rho_{[ij]}
             + 2F_{[i|k} \rho_{k|j]} + 2\kap_{[i}\vphi'_{j]}\\
  2\tho' \vphi_i + \eth_j F_{ji} - \eth_i F
         &=& (2\rho'_{[ij]}-\rho' \del_{ij}) \vphi_j
                      - 2F_{ij} \tau_j - 2 F\tau_i
                      + (2\rho_{(ij)}-\rho \del_{ij}) \vphi'_j \label{max3}\\
  \eth_{[i} F_{jk]} \label{max4}
         &=& \vphi_{[i} \rho'_{jk]} + \vphi'_{[i} \rho_{jk]}
\end{eqnarray}
A further three equations can be obtained by priming equations (\ref{max1}),(\ref{max2}) and (\ref{max3}).  We will often make use of the combination $ \del_{ij} (\ref{max1})- (\ref{max2}) $:
\begin{multline}\label{max1-2}
  \tho(F_{ij}+\del_{ij} F)
         = 2 \eth_{[i} \vphi_{j]} - \del_{ij}\eth_k \vphi_k - 2\tau'_{[i} \vphi_{j]} - 2F \rho_{[ij]}
             - 2F_{[i|k} \rho_{k|j]} - 2\kap_{[i}\vphi'_{j]} \\
           + \del_{ij}(\tau'_k \vphi_k + \rho_{kl} F_{kl} - \rho F - \kap_k \vphi'_k) 
\end{multline}
Now consider the combination $\tho(\ref{max3}) + \eth_j (\ref{max1-2})$.  This gives
\begin{eqnarray}
  0 &=& (2\tho' \tho + \eth_j \eth_j) \vphi_i + 2[\tho,\tho']\vphi_i - 
         [\tho,\eth_j] (F_{ij}+F\del_{ij}) + [\eth_i,\eth_j] \vphi_j \nn \\
    &&  + \tho \big(-(2\rho'_{[ij]}-\rho' \del_{ij}) \vphi_j
                      + 2(F_{ij}+F\del_{ij}) \tau_j
                      - (2\rho_{(ij)}-\rho \del_{ij}) \vphi'_j\big)\label{eqn:maxmaster}\\
    &&  + \eth_i \big(-\rho_{jk} F_{jk} + \rho F - \tau'_j \vphi_j 
                     + \kap_j \vphi'_j\big)+ \eth_j \big(2\tau'_{[i} \vphi_{j]} + 2F \rho_{[ij]}
        + 2F_{[i|k} \rho_{k|j]} + 2\kap_{[i}\vphi'_{j]}\big).\nn
\end{eqnarray}

This involves second derivatives of $\vphi$, as well as of the boost weight 0 quantities $F_{ij}$ and $F$. However, the latter occur in the form of a commutator $[\tho,\eth_j](F_{ij} + F\delta_{ij})$ and can therefore be eliminated.  Now we consider first derivatives of Maxwell components other than $\vphi$.  We need to eliminate these from the equation if it is to decouple.

First consider terms involving $\tho$:
\begin{itemize}
  \item $\tho$ acts on $F$ and $F_{ij}$ through the combination $\tho(F_{ij} + F\delta_{ij})$, which we eliminate using equation \eqref{max1-2}.

  \item Terms involving $\tho \vphi'_i$ are eliminated using equation \eqref{max3}$'$.

  \item Terms in which $\tho$ acts on $\rho_{ij}$, $\tau_i$ and $\rho'_{ij}$ are eliminated using the Newman-Penrose equations (\ref{NP1}), (\ref{NP2}) and (\ref{NP4})$'$ respectively (see Appendix \ref{app:ghpeqns}). 
  \end{itemize}
The resulting equation is very long:
\begin{multline}\label{eqn:maxfull}
  \Big[(2 \tho \thop + \eth_{j} \eth_{j} + \rhop \tho + \rho \thop - 2\taup_{j} \eth_{j}
        - 2 \tau_{j}\eth_{j}) \vphi_{i}  \\
       +( - \rhop_{ij} \tho - 2   \tau_{i} \eth_{j} + \rhop_{ji} \tho
      - \rho_{ij} \thop + [\eth_{i}, \eth_{j}]  + 2 \tau_{j}\eth_{i} - \rho_{ji} \thop )\vphi_{j} \\
       - \kap_{i} \kappap_{j} \vphi_{j} 
       - 2 \vphi_{j} \rho_{k i} \rhop_{jk} + \vphi_{j} \rho_{k j} \rhop_{i k} 
       - 2 \vphi_{j} \tau_{j} \taup_{i} + 2 \vphi_{i} \tau_{j} \taup_{j} 
       - \vphi_{j} \rho_{i k} \rhop_{jk}  + \vphi_{j} \rho\rhop_{ji} \\
       + 2 \vphi_{j} \tau_{i} \taup_{j} 
       + \kap_{j} \kappap_{i} \vphi_{j}
       - \kap_{j} \kappap_{j} \vphi_{i} - \vphi_{i} \rho_{k j} \rhop_{jk} - 2\Phia_{ij} \vphi_{j} - \Phi \vphi_{i} 
       -  \tfrac{d-2}{d-1}\Lambda \vphi_{i}
\Big] \\
+ \Big[ \kap_{j} \thop (F_{ij}+F\del_{ij}) +\rho_{ji}  \eth_{j} F - \rho_{k i}  \eth_{j} F_{jk} 
        + 2  \rho_{ij} \eth_{j} F  + \rho_{k j}\eth_{j} F_{i k} - \rho_{jk} \eth_{i} F_{jk}
        +  \rho_{jk}\eth_{j} F_{i k}  \\
       - F \eth_{j} \rho_{ji}  - F_{jk} \eth_{j} \rho_{k i} + F \eth_{j} \rho_{ij}  + F_{ij} \eth_{k} \rho_{jk}  
        + F \eth_{i} \rho   - F_{jk} \eth_{i} \rho_{jk}  
        + 2 F \thop \kap_{i} + 2 F_{ij} \thop \kap_{j} \\
        - 5 F \rho_{ij} \tau_{j} - 2 F \Psi_{i} - 4 F_{ij} \rho_{jk} \tau_{k} - F_{ij} \Psi_{j}
         - F_{jk} \kap_{i} \rhop_{jk} + F_{jk} \kap_{j} \rhop_{i k} - F_{jk} \Psi_{j k i} 
        - F_{ij} \kap_{k} \rhop_{jk} \\
        + F_{ij} \kap_{j} \rhop_{k k} + F \rho_{ji} \tau_{j} - 3 F_{jk} \rho_{ji} \tau_{k} 
        - F \rho \tau_{i} + 2 F_{jk} \rho_{jk} \tau_{i} - F_{jk} \rho_{ij} \tau_{k} + F_{ij} \rho \tau_{j}
\Big] \\
+ \Big[\kap_{j}  \eth_{i} \vphip_{j} + \kap_{i}\eth_{j} \vphip_{j}  - \kap_{j}  \eth_{j} \vphip_{i} 
       + 2 \rho_{k i} \rho_{jk} \vphip_{j}  + \kap_{j} \tau_{j} \vphip_{i} + \rho_{i k} \rho_{k j}\vphip_{j} \\
         + \vphip_{j} \rho_{i k} \rho_{jk} - \kap_{j} \tau_{i} \vphip_{j}- \kap_{i} \tau_{j} \vphip_{j}  
        - \rho_{ji} \rho\vphip_{j} - \rho_{jk} \rho_{k j} \vphip_{i} + 2 \Om_{ij} \vphip_{j} 
\Big]= 0
\end{multline}
The only terms above involving derivatives of Maxwell components other than $\vphi_{i}$ are of the (schematic) form $\kappa \thop F$, $\kap \eth \vphi'$ and $\rho \eth F$.  We need to eliminate all of these from our equations if we are to obtain a decoupled equation for $\vphi_i$.  Consider first the former two, which are
\begin{equation}
  \kappa_j \thop \left( F_{i j} + F \delta_{i j} \right) + 2\kap_j \eth_{[i}\vphi'_{j]} + \kap_i \eth_j\vphi'_j
        = 2\kappa_j \thop \left( F_{i j} + F \delta_{i j} \right) + \dots
\end{equation}
where we have used \eqref{max1-2}$'$ to eliminate the $\kap \eth \vphi'$ terms in favour of $\kappa \thop F$ and some other terms not involving derivatives.
 
Now, the Maxwell equations cannot be used to eliminate the terms of the form $\kappa \thop F$ without re-introducing 1-derivative terms of the form $\kappa \eth \vphi'$. Hence the only way in which the $\kappa \thop F$ terms can be eliminated is if $\kappa_i=0$, hence the vector field $\ell$ must be geodesic for decoupling to be possible. We assume henceforth that this is the case.

Now examine the $\rho \eth F$ terms above. These are:
\begin{equation} \label{rhoethF}
  \rho_{ji} \eth_{j}F - \rho_{ki} \eth_{j}F_{jk} + 2 \rho_{ij} \eth_{j}F + \rho_{kj} \eth_{j}F_{ik} 
  - \rho_{jk} \eth_{i}F_{jk} + \rho_{jk} \eth_{j}F_{ik}
\end{equation}
To achieve decoupling, we need to eliminate these terms from the equation without introducing any 1-derivative terms (unless the derivative acts on $\vphi$). It is convenient to decompose $\eth_i F_{jk}$ into parts that transform irreducibly under $SO(d-2)$:
\begin{equation}
 \eth_i F_{jk} = {\cal F}_{ijk} + \frac{2}{d-3} \delta_{i[j} \eth_{|l} F_{l|k]},
 \end{equation}
where ${\cal F}_{ijk}$ is traceless and can be decomposed further into objects transforming irreducibly according to the Young tableaux {\tiny\yng(1,1,1)} and {\tiny\yng(2,1)}.  The quantity $\eth_i F$, transforms in the same way as $\eth_j F_{ji}$, i.e. as a vector ({\tiny\yng(1)}) under $SO(d-2)$.  The latter can be eliminated in favour of the former using equation \eqref{max3}, which gives $\eth_j F_{j i} = \eth_i F + \ldots$, where the ellipsis denotes terms in which derivatives act only on $\vphi$. The contribution of the ``vector'' terms to \eqref{rhoethF} is then
\begin{equation}
\label{ethFterms}
 \frac{2}{d-3} \left( \rho_{ji} + (d-3) \rho_{ij} - \rho \delta_{ij} \right) \eth_j F
\end{equation}
We can substitute our decomposition of $\eth F$ into the Maxwell equations. There are no Maxwell equations that can be used to eliminate $\eth_i F$ without reintroducing new derivative terms of the form $\rho \tho \vphi'$. Hence the only way in which the Maxwell equation will decouple is if the expression in brackets in \eqref{ethFterms} vanishes. The symmetric and antisymmetric parts of the resulting equation give
\begin{equation}
\label{zeroshearrot}
 \sigma_{ij} = 0 = (d-4) \omega_{ij},
\end{equation}
where $\sigma$ and $\omega$ are the shear and rotation of $\ell$ respectively (i.e.\ they are the symmetric tracefree and antisymmetric parts of $\rho_{ij}$).  Hence a necessary condition for decoupling is that $\ell$ be shearfree and, for $d>4$, rotation free (and hence hypersurface orthogonal since $\ell$ is geodesic).  We now assume $d>4$, so we set $\sigma_{ij}=\omega_{ij}=0$ henceforth, and therefore have
\begin{equation}
  \rho_{ij} = \frac{\rho}{d-2} \delta_{ij}.
\end{equation}
A spacetime admitting a null geodesic congruence with vanishing rotation and shear is called a Robinson-Trautman spacetime if $\rho \ne 0$ and a Kundt spacetime if $\rho =  0$.  It was shown in Refs.\ \cite{Ricci,RobTraut} that an Einstein spacetime of either of these types is algebraically special, with the vector field $\ell$ aligned with the congruence being a multiple WAND.  Therefore we can take $\Omega_{ij} = \Psi_{ijk} = 0$. Note that (\ref{NP3}) now implies $\eth_i \rho=0$.

It is now guaranteed that we can use equation \eqref{max3} to eliminate ``vector terms'' of the form $\eth F$ from  \eqref{eqn:maxfull}. Upon doing so, we find that the terms involving ${\cal F}_{ijk}$ all drop out. The commutators $[\tho,\thop]$ and $[\eth_i, \eth_j]$ can be used to tidy up the equation, giving
\begin{multline}\label{maxwellexpanding}
  0 =  \left[2 \thop\tho + \eth_{j} \eth_{j} + \rhop \tho + \tfrac{d+2}{d-2} \rho \thop - 4 \tau_j \eth_{j}
         \right]\vphi_{i} 
          + 2( \tau_j \eth_{i} - \tau_i \eth_{j})\vphi_{j} \\
    + \left[ 3 \Phi_{i j} - \Phi_{j i} - \frac{2\rho}{d-2} \rhop_{[ij]}+  \left( \Phi  + \frac{\rho \rhop}{d-2}  
        - \frac{2d-3}{d-1} \Lambda  \right) \delta_{ij} \right] \vphi_j \\ 
    + \frac{d-4}{d-2} \rho \left[\tau_j \left( F_{ij} - F \delta_{ij} \right)  + \frac{\rho}{d-2} \vphip_i\right] .
\end{multline}
The only term involving $\vphi'_i$ is the final one, so for $\vphi'_i$ to decouple we need $(d-4) \rho = 0$. This also ensures that the terms involving $F_{ij}$ and $F$ drop out of the equation. Hence decoupling requires $\rho=0$ (since $d>4$), which implies
$\rho_{ij}=0$,  so $\ell$ must be free of expansion as well as shear and rotation.  That is, the spacetime must be Kundt.  The equation reduces to
\begin{multline}
  \left[2 \thop\tho + \eth_{j} \eth_{j} + \rhop \tho - 4 \tau_j \eth_{j} + \Phi - \frac{2d-3}{d-1} \Lambda 
         \right]\vphi_{i} 
          + ( 2\tau_j \eth_{i} - 2\tau_i \eth_{j} + 3 \Phi_{i j} - \Phi_{j i})\vphi_{j} = 0.
\end{multline}
which is equivalent to \eqref{eqn:maxpert}.

To summarize, for $d>4$, $\vphi_i$ satisfies a second-order decoupled equation if, and only if, $\ell$ is geodesic with vanishing expansion, rotation and shear, i.e., if, and only if, the spacetime is Kundt.

Note the presence of factors of $(d-4)$ in several of our equations above. When $d=4$, it is not necessary for the rotation $\om_{ij}$ of $\ell$ to vanish in equation \eqref{zeroshearrot}, or for the expansion $\rho$ to vanish in equation (\ref{maxwellexpanding}). Indeed, in 4d, all that is required is that $\ell$ be geodesic and shearfree, which is equivalent (by the Goldberg-Sachs theorem) to the spacetime being algebraically special. 

It is clear that $\vphi'_i$ will satisfy a second-order decoupled equation (the prime of the above equation) if, and only if, $\nb$ is geodesic with vanishing expansion, rotation and shear. Hence $\vphi_i$ and $\vphi'_i$ both will satisfy second order decoupled equations if, and only if, $\kappa_i = \kappa'_i = \rho_{ij} = \rho'_{ij} = 0$.  We call such a spacetime ``doubly Kundt''.

We have considered the conditions for decoupling of a Maxwell 2-form field. It would be interesting to consider the more general case of a $p$-form field with $p>2$. The equations of motion for such a field are given in GHP form in Ref. \cite{higherghp}.

\subsection{The Schwarzschild solution}
Consider the special case of the higher-dimensional Schwarzschild solution, which is not Kundt. This solution has $\rho_{ij} = \frac{\rho}{d-2} \delta_{ij}$ and $\tau_i=0$ (a consequence of spherical symmetry). The latter implies that the terms in $F_{ij}$ and $F$ drop out of equation \eqref{maxwellexpanding} leaving us with an equation of the form
\begin{equation}
 ( {\cal L} \vphi)_i +  \frac{(d-4)}{(d-2)^2} \rho^2 \vphi'_i = 0,
\end{equation}
where ${\cal L}$ is a second order differential operator.  The second term remains an obstruction to decoupling.  For the Schwarzschild solution, the two multiple WANDs have identical properties so we can take the prime of the equation to obtain
\begin{equation}
 ( {\cal L}' \vphi')_i +  \frac{(d-4)}{(d-2)^2} \rho'^2 \vphi_i = 0,
\end{equation}
and hence
\begin{equation}
  \left[  {\cal L}' \left( \frac{1}{\rho^2} {\cal L} \vphi \right) \right]_i - \frac{(d-4)^2}{(d-2)^4} \rho'^2 \vphi_i = 0.
\end{equation}
So in fact $\vphi_i$ does satisfy a decoupled equation but it is fourth order in derivatives. Note that we had to make use of several special properties of the Schwarzschild solution to obtain this result. It would be interesting to investigate more generally the circumstances under which one can obtain a decoupled equation of higher order for $\vphi_i$. 

\section{Decoupling of gravitational perturbations}\label{sec:gravdecoupling}

\subsection{Introduction and main result}

We now move on to gravitational perturbations.  In Lemma \ref{lem:Ominv} we found a set of gauge invariant quantities $\Om^{(1)}_{ij}$ under the assumption that $\lb$ was a multiple WAND of the background spacetime, so we shall consider gravitational perturbations of an algebraically special Einstein spacetime, for which we can take $\ell$ to be a multiple WAND.  Hence $\Omega_{ij}$ and $\Psi_{ijk}$ vanish in the background, so we can treat them as first order quantities: $\Omega_{ij} = \Omega^{(1)}_{ij}$, $\Psi_{ijk}=\Psi^{(1)}_{ijk}$. Therefore we shall not bother including a superscript ${}^{(1)}$ on $\Omega$ or $\Psi$ below.

The final result will be similar to that of the electromagnetic perturbations; we will find that we can only achieve decoupling when the spacetime is Kundt, that is when it admits a non-expanding, non-shearing, non-twisting null geodesic congruence.  We will show that gravitational perturbations of such a Kundt spacetime are described by
\begin{multline}\label{eqn:gravperts}
  \left(2\tho'\tho+ \eth_k \eth_k + \rho'\tho - 6\tau_k\eth_k  + 4\Phi - \tfrac{2d}{d-1} \La \right) \Om_{ij}\\
  + 4\left(\tau_k\eth_{(i|}- \tau_{(i|}\eth_k + \Phis_{(i|k} + 4\Phia_{(i|k}\right) \Om_{k|j)} 
  + 2\Phi_{ikjl} \Om_{kl} = 0,
\end{multline}
where all quantities except $\Omega$ are evaluated in the background geometry (e.g. $\Phi$ denotes $\Phi^{(0)}$ etc.)

In a doubly Kundt spacetime, $\Omega'_{ij}$ also will satisfy a decoupled equation, which is given by priming the above equation.

\subsection{Derivation of main result}

We follow as closely as possible the approach of Ref. \cite{stewperts}. We start by obtaining an equation in which second derivatives act only on $\Omega_{ij}$. Consider the equations
\begin{multline}\label{eqn:A}
  0   = - \eth_k \Om_{ij} - \del_{jk} \eth_l \Om_{il} + \eth_j \Om_{ik} -\tho (\Ps_i \del_{jk} + \Ps_{ijk}) 
          + \del_{jk}(\Phi_{li} - 2\Phi_{il} - \Phi \del_{il}) \kap_l  \\
        + (-2\Phi_{i[k|} \del_{j]l} + 2\del_{il} \Phia_{kj} + \Phi_{ilkj})\kap_l
          + \del_{jk} \left[ - \Psi_i \rho - \rho_{il} \Psi_l 
          - (\Psi_{mil}+\Psi_{iml})\rho_{lm} \right]\\
        + 2 (\Ps_{[k|} \del_{il} + \Ps_i\del_{[k|l} + \Ps_{i[k|l} + \Ps_{[k|il}) \rho_{l|j]}  
          + (\Om_{il} \tau'_l \del_{jk} - \Om_{ik} \tau'_{j} + \Om_{ij} \tau'_{k})
\end{multline}
and
\begin{multline}\label{eqn:B}
  0 = -2\tho' \Om_{ij} + \eth_k( \Psi_i \del_{jk} + \Psi_{ijk})
                    + \left( -\Om_{ij} \rho' + 2 \Om_{ik} \rho'_{[jk]} \right)
                    - 4(\Psi_{(i}\del_{j)k} + \Psi_{(ij)k})\tau_k \\
       + \Phi_{jk} \rho_{ik} - \Phi_{kj} \rho_{ik} + \Phi_{ik} \rho_{jk} - \Phi_{ki} \rho_{kj}
                   + 2 \Phi_{ik} \rho_{kj} - \Phi_{ij} \rho + \Phi_{ikjl} \rho_{kl} + \Phi \rho_{ij}.
\end{multline}
Equation \eqref{eqn:A} is obtained by taking various linear combinations and contractions of the Bianchi equations (\ref{B1}) (in Appendix \ref{app:ghpeqns}), while equation \eqref{eqn:B} is constructed from the symmetric part of (\ref{B2}) and a contraction of (\ref{B3}).  These equations are {\it exact}: no decomposition into background and perturbation has been performed at this stage.

Now we consider the linear combination $\eth_k \eqref{eqn:A}+\tho \eqref{eqn:B}$.
This contains second derivatives acting on $\Omega_{ij}$ and on $\Psi_{i j k}$. However, the point of taking this particular combination is that the second derivatives of $\Psi_{ijk}$ occur in the combination $- [\tho,\eth_k ] ( \Psi_{i j k} + \Psi_i \delta_{j k} )$ and therefore can be eliminated in favour of terms involving one or zero derivatives of $\Psi_{i j k}$ using the formula (\ref{C2}) for the commutator $[\tho,\eth_k]$.

We can also symmetrize the entire equation on $ij$ without losing any useful information, as the antisymmetric terms do not contain any second derivatives of $\Om$.  This reduces the equation to
\begin{eqnarray}
 0 &=& - (2\tho'\tho+ \eth_k\eth_k) \Om_{ij} - 2[\tho,\tho']\Om_{ij} - [\eth_{(i|},\eth_k]\Om_{k|j)}
       + \tho(\To_{ijkl} \rho'_{kl}) - \eth_l (\To_{ijkl}\tau'_k) \nn\\
   & & +[\tho, \eth_k]\Tp_{ijk} - 4\tho ( \Tp_{ijl}\tau_l ) 
       - 2\eth_{(i|} ( \Tp_{|j)lk}\rho_{kl}) + 2\eth_l(\Tp_{(j|lk} \rho_{k|i)}) - 2\eth_l(\Tp_{ijk}\rho_{kl})\nn\\
   & & +\tho(\Tf_{ikjl}\rho_{kl}) - \eth_l (\Tf_{ikjl}\kap_k) \label{eqn:star2}\label{master}
\end{eqnarray}
where
\begin{eqnarray}
  \Tf_{ikjl} & \equiv & \Phi_{(i|k|j)l} + \Phi \del_{(i|k}\del_{|j)l} - \Phis_{ij}\del_{kl} 
                        + (2\Phi_{(i|l}-\Phi_{l(i|})\del_{k|j)}
                        + (2\Phi_{(i|k}-\Phi_{k(i|})\del_{l|j)},\\
  \Tp_{ijk}  & \equiv & \Psi_{(ij)k} + \Psi_{(i}\del_{j)k},\\
  \To_{ijkl} & \equiv & -\Om_{ij}\del_{kl} + \Om_{(i|l}\del_{k|j)} - \Om_{(i|k} \del_{l|j)}.
\end{eqnarray}
Note that these quantities satisfy the following relations:
\begin{equation}\label{eqn:Tfid}
  \Tf_{ijkl} = \Tf_{(i|j|k)l} = \Tf_{i(j|k|l)}, \quad \Tf_{ijil} = 0 \eqand \Tf_{ijkj} = -(d-2)\Phis_{ik} + \Phi \del_{ik},
\end{equation}
\begin{equation}\label{eqn:Tpid}
  \Tp_{ijk} = \Tp_{(ij)k}, \quad \Tp_{iik} = 0 \eqand \Tp_{iji} = \half d\Psi_j
\end{equation}
\begin{equation}\label{eqn:Toid}
  \To_{ijkl} = \To_{(ij)kl},\quad 
  \To_{ij(kl)} = -\Om_{ij}\del_{kl} \quad 
  \To_{iikl} = 0 \eqand 
  \To_{ijkk} = -(d-2)\Om_{ij}.
\end{equation}
In this notation, the parts of \eqref{eqn:A} and \eqref{eqn:B} symmetric on $ij$ become
\begin{equation}\label{eqn:Asimp}
  \tho \Tp_{ijk} - \eth_l \To_{ijlk} = - \To_{ijlk}\tau'_l + 2\Tp_{(i|kl} \rho_{l|j)} 
                   - 2\Tp_{ijl}\rho_{lk} - 2\Tp_{l(i|m}\rho_{ml} \del_{k|j)} - \Tf_{ikjl}\kap_l
\end{equation}
and
\begin{equation}\label{eqn:Bsimp}
  -\eth_k \Tp_{ijk} + 2\tho'\Om_{ik} = \Tf_{ikjl}\rho_{jl} - 4\Tp_{ijk}\tau_k 
                                       + \To_{ijkl}\rho'_{kl}.
\end{equation}
Next we perform the following steps:
\begin{enumerate}
  \item Use the commutator (\ref{C2}) of Appendix \ref{app:ghpeqns} to eliminate the terms $[\tho, \eth_k]\Tp_{ijk}$ from \eqref{master} (note that this introduces a new kind of term, of the schematic form $\kap \tho' \Psi$).
  \item Expand out the brackets using the Leibniz rule for GHP derivatives.

  \item Eliminate the term $\tho\Tp_{ijk}$ using equation \eqref{eqn:Asimp}.

  \item Use the NP equations (\ref{NP1}), (\ref{NP2}) and (\ref{NP4})$'$ from Appendix \ref{app:ghpeqns} to eliminate terms 
in which $\tho$ acts on $\rho_{ij}$, $\tau_i$ and $\rho'_{ij}$ respectively.

  \item Take a linear combination of the Bianchi equations (\ref{B2},\ref{B3},\ref{B4}) to get an equation
  \begin{eqnarray}\label{eqn:thophi}
    \tho \Tf_{ikjl} &=& \tho' \To_{ijkl} + \eth_{(i|}\Psi_{l|j)k} - \eth_l\Psi_{(ij)k} 
                        - \del_{(i|k}\del_{|j)l}\eth_m \Psi_m
                        + \del_{kl}\eth_{(i}\Psi_{j)} \nn\\ 
                    & & + (-2\eth_l\Psi_{(i|} + \eth_{(i|} \Psi_l)\del_{k|j)}
                        + (-2\eth_k\Psi_{(i|} + \eth_{(i|} \Psi_k)\del_{l|j)} + \dots,
  \end{eqnarray}
  where the ellipsis indicates terms that involve no derivatives.  Use this to eliminate $\tho \Tf_{ikjl}$ from \eqref{master}.
  \item Use a combination of (\ref{B5}) and (\ref{B7}) to show that
      \begin{equation}
        \eth_l \Tf_{ijkl} = 3\tho'\Tp_{ijk} + 3\Tf_{ikjl}\tau_l + \dots
      \end{equation}
where the ellipsis denotes first order terms not involving any derivatives.  Use this to eliminate $\eth_l \Tf_{ijkl}$ from \eqref{master}.
\end{enumerate}
The resulting equation is very long so we shall not write it out in full. It has the schematic form
\begin{multline}\label{eqn:schematic}
  (\tho'\tho + \eth \cdot \eth + [\tho,\tho'] + [\eth,\eth] + \rho'\tho + \rho\tho' + \tau\eth + \tau'\eth 
        + \tau\tau' + \rho\rho' + \Phi ) \Om \\ 
      + \kap \tho' \Psi + \rho \eth \Psi + ( \tau \kap + \tau'\kap + \rho^2)\Phi 
      + ( \kap \rho )\Psi' + (\tau\rho + \tau'\rho + \kap \rho' + \tho'\kap + \eth\rho )\Psi =0
\end{multline}
Here, we neglect terms that are of quadratic order or higher when we decompose quantities into a ``background'' piece and a perturbation. Recall that $\Omega$ and $\Psi$ are first order quantities.
Note that the only terms containing derivatives of Weyl components other than $\Omega_{ij}$ are of the schematic form $\kappa \tho' \Psi$ and $\rho \eth \Psi$.  For decoupling to occur, these must vanish for any possible perturbation.  We shall now examine the circumstances under which we can eliminate these terms.

The detailed form of the $\kappa \tho' \Psi$ terms is
\begin{equation}
 4 \kappa_k \tho' ( \Psi_{(ij)k} + \Psi_{(i} \delta_{j) k})
\end{equation}
If $\kappa_i^{(0)} \ne 0$ then there is nothing we can do to eliminate these terms.  The only Bianchi equation containing $\tho'\Psi$ is \eqref{B5}, and using this again would reintroduce the 1-derivative terms that we have eliminated above.  Hence the only way for these terms to drop out is for $\kappa_i$ to vanish in the background.  Hence $\kappa_i^{(0)}=0$ is a necessary condition for decoupling.  Henceforth we assume $\kappa_i$ is a first-order quantity, in which case the above terms become second order terms and can be neglected.

Recall that $\kappa_i^{(0)}=0$ is equivalent to the statement that $\ell$ is geodesic in the background. This places no further restrictions on the spacetime, as it was shown in Ref.\ \cite{nongeo} that any Einstein spacetime admitting a multiple WAND also admits a {\it geodesic} multiple WAND.

Having set $\kappa_i^{(0)}=0$, the only remaining terms involving derivatives of Weyl components other than $\Omega$ are of the form $\rho \eth \Psi$.  The detailed form of these terms is:
\begin{multline} \label{eqn:rhoethPsi}
  4 \rho_{(i|k} \eth_k \Psi_{|j)} 
  + \rho_{kl} \left[ 2\eth_l \Psi_{(ij)k} + \eth_{(i|}\Psi_{l|j)k} - \eth_{(i}\Psi_{j)kl}
                     + 2\eth_k \Psi_{(ij)l} - \eth_{(i|} \Psi_{k|j)l} \right]\\
  + \rho_{k(i|} \left[ -\eth_l \Psi_{|j)lk} - \eth_l \Psi_{l|j)k} 
                       - \eth_{|j)} \Psi_k + 2\eth_k \Psi_{|j)} \right]
\end{multline}
For decoupling we need to eliminate these terms in favour of terms in which derivatives act only on $\Omega$.

Certain combinations of terms of the form $\eth \Psi$ can be eliminated using Bianchi equations. In order to understand precisely what kinds of terms can be so eliminated, we can decompose $\eth_i \Psi_{jkl}$ into parts that transform irreducibly under $SO(d-2)$.  If we do the same for the Bianchi equations at our disposal (or combinations of them such as \eqref{eqn:B}) then we will see which irreducible parts of $\eth \Psi$ can be eliminated from the above equation.

Decomposing into tracefree and trace parts gives, for $d>4$:
\begin{equation}
 \eth_i \Psi_{jkl} = V_{ijkl} + 2 \delta_{i[k|} W_{j|l]} + \delta_{ij} X_{kl} 
                     + 2 \delta_{j[k|} Y_{i|l]} + 2 \delta_{i[k|} \delta_{j|l]} Z,
\end{equation}
where $V_{ijkl}$ is traceless and satisfies $V_{i[jkl]}=V_{ij(kl)}=0$. The other terms are given by
\begin{equation}
 W_{[ij]} = \frac{1}{2} X_{ij} = \frac{1}{d(d-4)} \left( - (d-3) \eth_k \Psi_{[ij]k} + \eth_{[i} \Psi_{j]} \right), 
 \end{equation}
 \begin{equation}
 Y_{[ij]} 
 =  \frac{1}{d(d-4)} \left( 3 \eth_k \Psi_{[ij]k} -(d-1) \eth_{[i} \Psi_{j]} \right),
\end{equation}
\begin{equation}
 W_{(ij)} = \frac{1}{(d-2)(d-4)} \left( -(d-3) \eth_k \Psi_{(ij)k} + \eth_{(i} \Psi_{j)} - \eth_k \Psi_k \delta_{ij} \right),
\end{equation}
\begin{equation}
Y_{(ij)} = \frac{1}{(d-2)(d-4)} \left( \eth_k \Psi_{(ij)k} - (d-3)  \eth_{(i} \Psi_{j)} + \eth_k \Psi_k \delta_{ij} \right),
\end{equation}
\begin{equation}
 Z = \frac{1}{(d-2)(d-3)} \eth_k \Psi_k.
\end{equation}
Note that $W_{(ij)}$ and $Y_{(ij)}$ are traceless and $X_{(ij)}=0$.

The traceless part $V_{ijkl}$ can be decomposed further into parts that transform irreducibly under $SO(d-2)$. The relevant irreducible representations correspond to Young tableaux with 4 boxes. However, it turns out that we will not need to discuss these. As well as these quantities, we have two independent quantities transforming as {\tiny$\yng(1,1)$}, namely $W_{[ij]}$ and $Y_{[ij]}$, two quantities transforming as {\tiny$\yng(2)$}, namely $W_{(ij)}$ and $Y_{(ij)}$, and a singlet $Z$.

Consider first the singlet $Z$, i.e., the trace $\eth_k \Psi_k$. The contribution of this to equation \eqref{eqn:rhoethPsi} is 
\begin{equation}
 4 (d-3) \sigma_{ij} Z = \frac{4}{d-2} \sig_{ij}  \eth_k \Psi_k,
\end{equation}
where the shear $\sigma_{ij}$ is the traceless symmetric part of $\rho_{ij}$.  In order to achieve decoupling, we would need to add to \eqref{eqn:rhoethPsi} a combination of Bianchi components containing a singlet term that cancelled this, and did not introduce any 1-derivative terms (e.g. $\tho \Phi$ terms) that we have already eliminated. However, 
there is no such combination.  For example, the singlet drops out of equation \eqref{eqn:B}. Therefore, the only way to eliminate the singlet term from our equation, as required for decoupling, is to set $\sigma_{ij}^{(0)}=0$, i.e., in the background geometry, the shear of the multiple WAND $\ell$ must vanish.  Henceforth we assume that this is the case.
 
Next consider the traceless symmetric tensors that arise in the above decomposition of $\eth_i \Psi_{jkl}$, i.e., $W_{(ij)}$ and $Y_{(ij)}$. The contribution of these to \eqref{eqn:rhoethPsi} is:
\begin{equation}\label{eqn:WY}
 - 5 \rho (W_{(ij)} + Y_{(ij)} ) 
 + \tfrac{1}{2}(d-10) \left(W_{(ik)} \om_{jk} + W_{(jk)} \om_{ik} \right)
 -\tfrac{3}{2}(d-2)\left( Y_{(ik)} \omega_{jk} + Y_{(jk)} \omega_{ik} \right),
\end{equation}
where $\omega_{ij} \equiv \rho_{[ij]}$.  

Now consider the Bianchi equations. The only combination of equations involving $W_{(ij)}$ and $Y_{(ij)}$ that does not introduce any 1-derivative terms that we have already eliminated is \eqref{eqn:Bsimp}, which gives an expression for 
\begin{equation}
  \eth_k \Tp_{ijk} \equiv -(d-2)\left(W_{(ij)}+Y_{(ij)}\right).
\end{equation}
We can use this to eliminate, say, $Y_{(ij)}$ from \eqref{eqn:WY}, i.e., we have $Y_{(ij)} = - W_{(ij)} + \ldots$, where the ellipsis denotes terms in which derivatives act only on $\Omega$. \eqref{eqn:WY} then reduces to
\begin{equation}
 2 ( d - 4) \left( W_{(ik)} \omega_{jk} + W_{(jk)} \omega_{ik} \right) + \ldots.
\end{equation}
Since we have no independent equation that will allow us to eliminate $W_{(ij)}$, we conclude that in order for the $\eth\Psi$ terms to decouple we must have $\omega_{ij}=0$ in the background, i.e., the multiple WAND $\ell$ must be free of both shear {\it and} rotation.  Note the factor of $d-4$: for $d=4$, vanishing rotation is {\it not} necessary for decoupling.\footnote{For $d=4$, $\Psi_{ijk} = -2 \delta_{i[j} \Psi_{k]}$, so the irreducible parts of $\eth_{i} \Psi_{jkl}$ are just the trace, tracefree symmetric and antisymmetric parts of $\eth_i \Psi_j$. Considering the trace gives $\sigma_{ij}=0$ as for $d>4$. The tracefree symmetric part can be eliminated with \eqref{eqn:B}. The antisymmetric part simply drops out of \eqref{eqn:rhoethPsi}, using the fact that all $2 \times 2$ antisymmetric matrices commute.}

Having set $\sigma^{(0)}_{ij} = \omega^{(0)}_{ij} = 0$, we find that that the 1-derivative terms \eqref{eqn:rhoethPsi} reduce to
\begin{equation}
 \frac{5 \rho}{d-2} \left( \eth_k \Psi_{(ij)k} + \eth_{(i} \Psi_{j)} \right)
    =  \frac{5 \rho}{d-2} \eth_k \Tp_{ijk}
\end{equation}
These terms can be eliminated from \eqref{eqn:schematic} with equation \eqref{eqn:Bsimp}.

In the resulting equation, we now use (\ref{NP3}) to argue that $\eth_i \rho$ is a first order quantity. It appears only when multiplied by $\Psi$, so such terms are second order and can be dropped. The only Weyl components that are now acted on by derivatives are $\Omega_{ij}$, and the equation has been reduced to the schematic form
\begin{equation}\label{eqn:schematic2}
  (\tho'\tho + \eth\cdot \eth + [\tho,\tho'] + [\eth,\eth] + \rho'\tho + \rho\tho' + \tau\eth + \tau'\eth 
        + \rho\rho' + \tau\tau' + \Phi ) \Om + \rho^2 \Phi + \tau\rho\Psi=0
\end{equation}
At this point, we can also simplify the form of the terms involving $\Om$, by using the commutators (\ref{C1},\ref{C3}) from Appendix \ref{app:ghpeqns} to eliminate the terms of the form $[\tho,\tho']\Om$ and $[\eth,\eth]\Om$ respectively, in favour of terms that involve at most first derivatives of $\Om$. 

The terms of the form $\Phi\rho$ are simplified by noting that equation evaluated \eqref{eqn:B} in the background geometry implies that
\begin{equation}
\label{Phiback}
 \rho^{(0)} \Phi^{(0)}_{ij} = \frac{1}{d-2} \rho^{(0)} \Phi^{(0)} \delta_{ij}.
\end{equation}
Equation \eqref{eqn:schematic2} now reduces to something sufficiently simple to write out explicitly: \begin{multline}\label{eqn:expstar}
     \left(2\tho'\tho+ \eth_k \eth_k + \rho'\tho + \tfrac{d+6}{d-2}\rho\tho' + \tfrac{2}{d-2}\rho\rho' 
           - 6\tau_k\eth_k + 4\Phi - \tfrac{2d}{d-1} \La \right) \Om_{ij}\\
    + \left(4\tau_k\eth_{(i|}- 4\tau_{(i|}\eth_k + \tfrac{2}{d-2}\rho(\rho'_{k(i|} - \rho'_{(i|k})
             +  4\Phis_{(i|k} + 16\Phia_{(i|k}\right) \Om_{k|j)} 
    + 2\Phi_{ikjl} \Om_{kl} \\
     + \frac{2\rho^2}{d-2}\left(\Phis_{ij} - \tfrac{1}{d-2}\Phi\del_{ij}\right)
       + 2\rho \tau_k  \left(\Psi_{(ij)k} - \Psi_{(i}\del_{j)k}
                                               +\tfrac{2}{d-2}\del_{ij}\Psi_k \right)=0. 
\end{multline}
This equation is the analogue of equation \eqref{maxwellexpanding} for the Maxwell field. Note that \eqref{Phiback} implies that $\rho (\Phi^S_{ij} - \tfrac{1}{d-2} \Phi \delta_{ij} )$ is a first order quantity.
To achieve decoupling we have to eliminate the terms not involving $\Om_{ij}$, i.e., those on the final line of this equation.  For $d=4$, this is automatic since the particular combination of $\Phi$ terms appearing in this equation vanishes identically (i.e. $\Phi^S_{ij} = \frac{1}{2} \Phi \delta_{ij}$ if $d=4$), as does the particular combination of $\Psi$ terms. For $d>4$, the only way of eliminating the $\Phi$ terms above is to set $\rho^{(0)}=0$, i.e., take $\rho$ to be first order. All terms on the final line above are then higher order and can be neglected.

Hence we see that, for $d>4$, decoupling requires that
\begin{equation}
 \kappa^{(0)}_i = \rho^{(0)}_{ij} = 0,
\end{equation}
i.e., the multiple WAND must be geodesic and free of expansion, rotation and shear. In other words, the spacetime must be Kundt. This is a necessary condition for decoupling; it is also sufficient since we now have an equation in which the only perturbed Weyl components that appear are $\Omega_{ij}$. 

The resulting decoupled equation is:
\begin{multline}
    \left(2\tho'\tho+ \eth_k \eth_k + \rho'\tho - 6\tau_k\eth_k  + 4\Phi - \tfrac{2d}{d-1} \La \right) \Om_{ij}\\
    + 4\left(\tau_k\eth_{(i|}- \tau_{(i|}\eth_k + \Phis_{(i|k} + 4\Phia_{(i|k}\right) \Om_{k|j)} 
    + 2\Phi_{ikjl} \Om_{kl} = 0.
\end{multline}
We remind the reader that $\Omega_{ij}$ is a first order quantity, so quantities multiplying $\Omega$ (e.g.\ $\Phi$, $\tau$) must be evaluated in the background geometry.

\subsection{Comment on the expanding case}

Just as we did for Maxwell perturbations, it is interesting to consider what happens if $\ell$ is geodesic with vanishing rotation and shear, but non-vanishing expansion (i.e. the spacetime is Robinson-Trautman). Under these circumstances, we have equation \eqref{eqn:expstar}, a perturbation equation for a gauge invariant quantity $\Om_{ij}$.  However, it contains two terms that obstruct the decoupling of the equation.  It is interesting to ask how these terms are consistent with gauge invariance. The answer is supplied by:
\begin{lemma}\label{lem:otherinvs}
  Let $\lb$ be an expanding, non-twisting, non-shearing geodesic multiple WAND for an Einstein spacetime of dimension $d>4$.  Then
  \begin{equation}\label{eqn:gaugeinv1}
    \left(\Phis_{ij} - \tfrac{1}{d-2}\Phi\del_{ij}\right)^{(1)}
  \end{equation}
  is a gauge invariant quantity. If $\tau_i^{(0)} \neq 0$, then
  \begin{equation}\label{eqn:gaugeinv2}
    \Psi^{(1)}_i \eqand \tau^{(0)}_k \Psi^{(1)}_{ijk}
  \end{equation}
 also are gauge invariant quantities.
\end{lemma}
The Schwarzschild black hole in arbitrary dimension is an example of a spacetime admitting such a multiple WAND (although in this case, $\tau^{(0)}_i=0$).  In four dimensions, \eqref{eqn:gaugeinv1} vanishes identically in all spacetimes, while the quantities \eqref{eqn:gaugeinv2} are not gauge invariant. 
\proof
From equation \eqref{Phiback} we have
\begin{equation}
\label{Phiback2}
\Phi_{ij}^{(0)} = \tfrac{1}{d-2}\Phi^{(0)}\del_{ij}
\end{equation}
in any such spacetime.  Hence we see immediately that \eqref{eqn:gaugeinv1} is invariant under infinitesimal coordinate transformations, and also under infinitesimal spins. Furthermore, Ref. \cite{RobTraut} showed that all such spacetimes are of algebraic Type D so we can choose our basis so that all Weyl tensor components with non-zero boost weight vanish.  Under an infinitesimal null rotation about $\lb$, equation (2.32,\cite{higherghp}) implies that, to first order in $z_i$,
\begin{equation}
   \Phis_{ij} \mapsto \Phis_{ij}+z_{(i}\Psi_{j)} - z_k\Psi_{(ij)k}, 
\end{equation}
but $\Psi$ is a first order quantity and hence $\Phi_{ij}^{S(1)}$ and $\Phi=\Phi_{ii}^{(1)}$ both are invariant in a Type D background.  An identical argument applies to null rotations about $\nb$, and hence \eqref{eqn:gaugeinv1} is a gauge invariant quantity.

For an algebraically special spacetime, $\Psi_{ijk}$ and $\Psi_i$ both vanish in the background, and so, to first order, they are invariant under infinitesimal spins and infinitesimal coordinate transformations.  They are also invariant under infinitesimal null rotations about $\lb$, as these can only introduce terms involving $\Om_{ij}$ which also vanishes in the background.  We now consider the effect of an infinitesimal null rotation about $\nb$.  Taking the prime of (2.34,\cite{higherghp}) implies that, to linear order,
\begin{equation}\label{eqn:psirot1}
 \Psi_{ijk}  \mapsto \Psi_{ijk} + \tfrac{2}{d-2} \Phi^{(0)} \delta_{i[j} z_{k]}  + z_l \Phi_{lijk}^{(0)}
\end{equation}
and 
\begin{equation}\label{eqn:psirot2}
 \Psi_{j}  \mapsto \Psi_{j} - \tfrac{d-1}{d-2} \Phi^{(0)} z_j,
\end{equation}
where we have used \eqref{Phiback2}. We will show that the quantities \eqref{eqn:gaugeinv2} are invariant under this transformation if $\tau_i^{(0)} \ne 0$.

Take a double trace of the Bianchi equation (\ref{B7}) for the background spacetimes.  This implies that $(d-4)\eth_k\Phi^{(0)} = 0$, and hence, for $d>4$, $\eth_k\Phi^{(0)} = 0$. The trace of (\ref{B5}) gives
\begin{equation}\label{eqn:ethphi}
  \eth_j\Phi^{(0)} = \tfrac{d-1}{d-3} \tau_j^{(0)} \Phi^{(0)} ,
\end{equation}
and hence $\Phi^{(0)}=0$ if $\tau_i^{(0)} \ne 0$. From \eqref{Phiback2} we then have $\Phi_{ij}^{(0)} = 0$. Putting these results back into (B5) implies that $\Phi_{ijkl}^{(0)}\tau_l^{(0)} = 0$.
Inserting these results into (\ref{eqn:psirot1},\ref{eqn:psirot2}) implies that, although $\Psi_{ijk}$ is not invariant under infinitesimal null rotations about $\nb$, both $\tau_k \Psi_{ijk}$ and $\Psi_i$ are invariant, and hence both of these are new gauge invariant quantities, provided that $d>4$ and $\tau_i^{(0)} \neq 0$.\, $\Box$

\section{Discussion}\label{sec:discussion}

We have shown that, for linearized gravitational perturbations of an algebraically special spacetime, there exist local quantities $\Omega^{(1)}_{ij}$, linear in the perturbation, that are invariant under infinitesimal coordinate transformations and infinitesimal changes of basis. For perturbations of a type D background, e.g. a Myers-Perry black hole, both $\Omega^{(1)}_{ij}$ and $\Omega_{ij}^{'(1)}$ are gauge invariant. We emphasize that, irrespective of decoupling, the locality and gauge invariance of these quantities should make them useful in studies of gravitational perturbations.

We have shown that $\Omega^{(1)}_{ij}$ satisfies a decoupled equation of motion in $d>4$ dimensions if, and only if, the background is Kundt, i.e., admits a null geodesic congruence with vanishing expansion, rotation and shear. Therefore the decoupling property which is satisfied in the Kerr spacetime does not extend to the Myers-Perry spacetimes dimensions in an obvious way.  

When decoupling does occur, an important question is whether a solution of the decoupled equation uniquely characterizes the gravitational perturbation. If one has two solutions with the same $\Omega^{(1)}_{ij}$ then do they describe the same metric perturbation? This is equivalent to the question of whether there exist non-trivial linearized gravitational perturbations with $\Omega^{(1)}_{ij}=0$.

For perturbations of a Kerr black hole, this problem was addressed in Ref.\ \cite{waldtypeDpert}, where it was shown that a well-behaved solution with $\Psi^{(1)}_0=0$ must also have $\Psi^{(1)}_4=0$. A null rotation about $n$ can then be used to set $\Psi^{(1)}_1=0$ and a null rotation about $\ell$ can be used to set $\Psi^{(1)}_3=0$. It follows that the perturbation must preserve the type D condition to linear order. Since type D solutions are specified by just a few constants \cite{Kinnersley}, it is natural to expect that there will be only a finite number of solutions satisfying these conditions. Ref.\ \cite{waldtypeDpert} showed that the only well-behaved solutions correspond simply to perturbations in the mass or angular momentum of the Kerr solution.

For $d>4$, even if one can show that $\Omega^{(1)}_{ij} = 0$ implies that $\Omega_{ij}^{'(1)}=0$ then it is no longer true that one can use null rotations to set $\Psi^{(1)}_{ijk} = \Psi^{'(1)}_{ijk} = 0$. This is because a null rotation about $n$ contains fewer parameters than the number of independent components of $\Psi^{(1)}_{ijk}$ (whereas for $d=4$ both have 2 degrees of freedom). So for $d>4$ it seems likely that the perturbations overlooked by our decoupled equation are more general than perturbations preserving the type D condition. Nevertheless, since $\Omega^{(1)}_{ij}$ has the same number of degrees of freedom as the gravitational field, it seems reasonable to expect that our decoupled equation of motion captures ``nearly all'' of the information about linearized metric perturbations, i.e., that only a few special solutions are missed.
 
Sometimes it might not be enough to know the solution for $\Omega^{(1)}_{ij}$, one might need to know the metric perturbation explicitly. Ref.\ \cite{Wald:1978vm} gives a systematic procedure for constructing solutions of the linearized Einstein equation (in a certain gauge), given the existence of a decoupled equation of motion for a quantity linear in the metric perturbation. It seems likely that this procedure can be applied in the present case to generate solutions of the higher-dimensional linearized Einstein equation whenever $\Omega^{(1)}_{ij}$ satisfies a decoupled equation, i.e., in a Kundt background. 

Finally, we discuss an application of our decoupled equation. Any extreme black hole admits a near-horizon geometry \cite{Reall:2002bh}. It turns out that all known extreme vacuum black hole solutions have near-horizon geometries which take the form of a fibration over $AdS_2$ \cite{ads2}:
\begin{equation}
 ds^2 = L(y)^2 \left( - R^2 dT^2 + \frac{dR^2}{R^2} \right) 
        + g_{IJ}(y) \left( d\phi^I - k^I R dT \right) \left( d\phi^J - k^J R dT \right) + g_{MN}(y) dy^M dy^N,
\end{equation}
where $\partial/\partial \phi^I$, $I=1, \ldots, n$ are rotational Killing vector fields of the black hole and $k^I$ are constants. The metric in the first set of round brackets is the metric of $AdS_2$.
The metric depends non-trivially only on the $d-n-2$ coordinates $y^M$.
A calculation reveals that the vector fields dual to $-dT \pm dR/R^2$ are tangent to affinely parameterized null geodesics with vanishing expansion, rotation and shear. Hence the above spacetime is doubly Kundt so our decoupled equations can be used to study gravitational perturbations of it. This is under investigation \cite{nearhorizon}.

\section*{Acknowledgments}
Many of the calculations in this paper were checked using the computer algebra software Cadabra \cite{Cadabra}. HSR is a Royal Society University Research Fellow.  MND is supported by STFC.

\appendix

\section{Useful GHP equations}\label{app:ghpeqns}

Many of the results in Sections \ref{sec:maxdecoupling} and \ref{sec:gravdecoupling} were based on the Newman-Penrose equations, Bianchi equations and derivative commutators for the higher-dimensional Geroch-Held-Penrose formalism defined in Ref.\ \cite{higherghp}.  Here we reproduce these equations for convenience, in the case of an Einstein spacetime satisfying $R_{\mu\nu} = \La g_{\mu\nu}$.

\subsection{Newman-Penrose equations}
\newcounter{oldeq}
\setcounter{oldeq}{\value{equation}}
\renewcommand{\theequation}{NP\arabic{equation}}
\setcounter{equation}{0}
\begin{eqnarray}\label{NP1}
  \tho \rho_{ij} - \eth_j \kap_i &=& - \rho_{ik} \rho_{kj} -\kap_i \tau'_j - \tau_i \kap_j - \Om_{ij},\\[3mm]
  \tho \tau_i - \tho' \kap_i &=& \rho_{ij}(-\tau_j + \tau'_j) - \Psi_i,\label{NP2}\\[3mm]
  2\eth_{[j|} \rho_{i|k]}     &=& 2\tau_i \rho_{[jk]} + 2\kap_i \rho'_{[jk]} - \Psi_{ijk} ,\label{R:ethrho}\label{NP3}\\[3mm]
  \tho' \rho_{ij} - \eth_j \tau_i &=& - \tau_i \tau_j - \kap_i \kap'_j - \rho_{ik}\rho'_{kj}-\Phi_{ij}
                                      - \tfrac{\La}{d-1}\del_{ij}.\label{NP4}
\end{eqnarray}
Another four equations can be obtained by taking the prime $'$ of these four (i.e.\ by exchanging the vectors $\lb$ and $\nb$).
\renewcommand{\theequation}{A.\arabic{equation}}
\setcounter{equation}{\value{oldeq}}

\subsection{Bianchi equations}\label{sec:bianchi}
\setcounter{oldeq}{\value{equation}}
\renewcommand{\theequation}{B\arabic{equation}}
\setcounter{equation}{0}
{\noindent\bf Boost weight +2:}
\begin{eqnarray}
  \tho \Ps_{ijk} - 2 \eth_{[j}\Om_{k]i} 
                  &=& (2\Phi_{i[j|} \del_{k]l} - 2\del_{il} \Phia_{jk}-\Phi_{iljk})\kap_l \nn\\
                  && -2 (\Ps_{[j|} \del_{il} + \Ps_i\del_{[j|l} + \Ps_{i[j|l} 
                     + \Ps_{[j|il}) \rho_{l|k]} + 2 \Om_{i[j} \tau'_{k]},\label{B1}
\end{eqnarray}
{\bf Boost weight +1:}
\begin{eqnarray}
  - \tho \Phi_{ij} - \eth_{j}\Ps_i + \tho' \Om_{ij} 
                 &=& - (\Ps'_j \del_{ik} - \Ps'_{jik}) \kap_k + (\Phi_{ik} + 2\Phia_{ik} + \Phi \del_{ik}) \rho_{kj} 
                      \nonumber\\
                 &&  + (\Ps_{ijk}-\Ps_i\del_{jk}) \tau'_k - 2(\Ps_{(i}\del_{j)k} + \Ps_{(ij)k}) \tau_k 
                     - \Om_{ik} \rho'_{kj}, \label{B2}\\[3mm]
  -\tho \Phi_{ijkl} + 2 \eth_{[k}\Ps_{l]ij}
                 &=& - 2 \Ps'_{[i|kl} \kap_{|j]} - 2 \Ps'_{[k|ij}\kap_{|l]}\nn\\
                 &&  + 4\Phia_{ij} \rho_{[kl]} -2\Phi_{[k|i}\rho_{j|l]} 
                     + 2\Phi_{[k|j}\rho_{i|l]} + 2 \Phi_{ij[k|m}\rho_{m|l]}\nn\\
                 &&  -2\Ps_{[i|kl}\tau'_{|j]} - 2\Ps_{[k|ij} \tau'_{|l]}
                     - 2\Om_{i[k|} \rho'_{j|l]} + 2\Om_{j[k} \rho'_{i|l]},
                     \label{B3}\\[3mm]
  -\eth_{[j|} \Ps_{i|kl]}
                 &=& 2\Phia_{[jk|} \rho_{i|l]} - 2\Phi_{i[j} \rho_{kl]} 
                     + \Phi_{im[jk|} \rho_{m|l]} - 2\Om_{i[j} \rho'_{kl]},\label{B4}
\end{eqnarray}
{\bf Boost weight 0:}
\begin{eqnarray}
  \tho' \Ps_{ijk} -2 \eth_{[j|}\Phi_{i|k]} 
                 &=& 2(\Ps'_{[j|} \del_{il} - \Ps'_{[j|il}) \rho_{l|k]}
                     + (2 \Phi_{i[j}\del_{k]l} - 2\del_{il}\Phia_{jk} - \Phi_{iljk}) \tau_l \nn\\
                 &&  + 2 (\Ps_i \del_{[j|l} -  \Ps_{i[j|l})\rho'_{l|k]} + 2\Om_{i[j}\kap'_{k]},
                     \label{B5}\\[3mm]
  -2\eth_{[i} \Phia_{jk]} 
                 &=& 2\Ps'_{[i} \rho_{jk]} + \Ps'_{l[ij|} \rho_{l|k]} 
                     - 2\Ps_{[i} \rho'_{jk]} - \Ps_{l[ij|} \rho'_{l|k]},\label{B6}\\[3mm]
  -\eth_{[k|} \Phi_{ij|lm]} 
                 &=& - \Ps'_{i[kl|} \rho_{j|m]} + \Ps'_{j[kl|} \rho_{i|m]} 
                     - 2\Ps'_{[k|ij} \rho_{|lm]}\nn\\
                  && - \Ps_{i[kl|} \rho'_{j|m]} + \Ps_{j[kl|} \rho'_{i|m]} 
                     - 2\Ps_{[k|ij} \rho'_{|lm]}.\label{B7}
\end{eqnarray}
Another five equations are obtained by applying the prime operator to equations (\ref{B1})-(\ref{B5}) above.
\renewcommand{\theequation}{A.\arabic{equation}}
\setcounter{equation}{\value{oldeq}}

\subsection{Commutators of derivatives}
\setcounter{oldeq}{\value{equation}}
\renewcommand{\theequation}{C\arabic{equation}}
\setcounter{equation}{0}
Acting on a GHP scalar of boost weight $b$ and spin $s$, commutators of GHP derivatives can be simplified by:
\begin{eqnarray}
[\tho, \tho']T_{i_1...i_s} 
         &=& (-\tau_j + \tau'_j) \eth_jT_{i_1...i_s} + 
                    b\left( -\tau_j\tau'_j + \kap_j\kap'_j + \Phi \right)T_{i_1...i_s} \nn\\
         &&  + \sum_{r=1}^s \left(\kap_{i_r} \kap'_{j} - \kap'_{i_r} \kap_{j} 
                                  + \tau'_{i_r} \tau_{j} - \tau_{i_r} \tau'_{j} + 2\Phia_{i_r j}
                            \right) T_{i_1...j...i_s}, \label{comm:thotho}\label{C1}\\[3mm]
[\tho, \eth_i]T_{k_1...k_s}
         &=& -(\kap_i \tho' + \tau'_i\tho +\rho_{ji}\eth_j)T_{k_1...k_s}
             + b\left(-\tau'_j\rho_{ji} + \kap_j\rho'_{ji} + \Psi_i \right)T_{k_1...k_s} \nn\\
         &+&  \sum_{r=1}^s \left( \kap_{k_r}\rho'_{li} - \rho_{k_r i}\tau'_l
            + \tau'_{k_r} \rho_{li} - \rho'_{k_r i} \kap_l - \Psi_{ilk_r}\right) T_{k_1...l...k_s},
            \label{comm:thoeth}\label{C2}\\[3mm]
[\eth_i,\eth_j]T_{k_1...k_s}
         &=& \left(2\rho_{[ij]} \tho' + 2\rho'_{[ij]} \tho \right) T_{k_1...k_s}
                   + b \left(2\rho_{l[i|} \rho'_{l|j]} + 2\Phia_{ij}\right) T_{k_1...k_s}\nn\\
         && + \sum_{r=1}^s \left(2\rho_{k_r [i|} \rho'_{l|j]} + 2\rho'_{k_r [i|} \rho_{l|j]} 
                                + \Phi_{ijk_r l} + \tfrac{2\La}{d-1} \del_{[i|k_r}\del_{|j]l} 
                           \right) T_{k_1...l...k_s}. \label{comm:etheth} \label{C3}
\end{eqnarray}
The result for $[\tho', \eth_i]$ can be obtained from \eqref{comm:thoeth}$'$. 
\renewcommand{\theequation}{A.\arabic{equation}}
\setcounter{equation}{\value{oldeq}}

\end{document}